Original Paper

# The Relationship between Deteriorating Mental Health Conditions and Longitudinal Behavioral Changes in Google and YouTube Usages among College Students in the United States during COVID-19: Observational Study

[1]Anis Zaman*, [1]Boyu Zhang*, [1]Ehsan Hoque, [2]Vincent Silenzio, [1]Henry Kautz

[1]Department of Computer Science, University of Rochester, Rochester, NY, USA

[2]Department of Urban-Global Public Health, Rutgers University, Jersey City, NJ, USA

*Equal Contribution

Correspondence to

Mr. Boyu Zhang

Department of Computer Science, University of Rochester, Rochester, NY 14627, USA

E-mail: bzhang25@u.rochester.edu

&

Mr. Anis Zaman

Department of Computer Science, University of Rochester, Rochester, NY 14627, USA

E-mail: azaman2@cs.rochester.edu

## Abstract

**Background:** Mental health problems among the global population are worsened during the coronavirus disease (COVID-19). Yet, current methods for screening mental health issues rely on in-person interviews, which can be expensive, time-consuming, blocked by social stigmas and quarantines. Meanwhile, how individuals engage with online platforms such as Google Search and YouTube undergoes drastic shifts due to COVID-19 and subsequent lockdowns. Such ubiquitous daily behaviors on online platforms have the potential to capture and correlate with clinically alarming deteriorations in mental health profiles of users in a non-invasive manner.

**Objective:** The goal of this study is to examine, among college students in the United States, the relationship between deteriorating mental health conditions and changes in user behaviors when engaging with Google Search and YouTube during COVID-19.

**Methods:** This study recruited a cohort of undergraduate students (N=49) from a U.S. college campus during January 2020 (prior to the pandemic) and measured the anxiety and depression levels of each participant. The anxiety level was assessed via the General Anxiety Disorder-7 (GAD-7). The depression level was assessed via the Patient Health Questionnaire-9 (PHQ-9). This study followed up with the same cohort during May 2020 (during the pandemic), and the anxiety and depression levels were assessed again. The longitudinal Google Search and YouTube history data of all participants were anonymized and collected. From individual-level Google Search and YouTube histories, we developed 5 signals that can quantify shifts in online behaviors during the pandemic. We then assessed the differences between groups with and without deteriorating mental health profiles in terms of these features.

**Results:** Of the 49 participants, 41% (n=20) of them reported a significant increase (increase in the PHQ-9 score ≥ 5) in depression, denoted as DEP; 45% (n=22) of them reported a significant increase (increase in the GAD-7 score ≥ 5) in anxiety, denoted as ANX. Of the 5 features proposed to quantify online behavior changes, statistical significances were found between the DEP and non-DEP groups for all of them ($P \leq .01$, effect sizes $\eta^2_{partial}$ ranging between 0.130 to 0.320); statistical significances were found between the ANX and non-ANX groups for 4 of them ($P \leq .02$, effect sizes $\eta^2_{partial}$ ranging between 0.115 to 0.231). Significant features included late-night online activities, continuous usages and time away from the internet, porn consumptions, and keywords associated with negative emotions, social activities, and personal affairs.

**Conclusions:** The results suggested strong discrepancies between college student groups with and without deteriorating mental health conditions in terms of behavioral changes in Google Search and YouTube usages during the COVID-19. Though further studies are required, our results demonstrated the feasibility of utilizing pervasive online data to establish non-invasive surveillance systems for mental health conditions that bypasses many disadvantages of existing screening methods.

## Introduction

### Background
Globally, mental health problems such as depression, anxiety, and suicide ideations are severely worsened during the coronavirus disease (COVID-19) [1–3], specifically for college students [4,5–7]. Yet, current methods for screening mental health issues and identifying vulnerable individuals rely on in-person interviews. Such assessments can be expensive, time-consuming, and blocked by social stigmas, not to mention the reluctancy induced by travel restrictions and exposure risks. It has been reported that very few patients in need were correctly identified and received proper mental health treatments on time under the current healthcare system [8,9]. Even with emerging Telehealth technologies and online surveys, the screening requires patients to actively reach out to care providers.

At the same time, because of the lockdown enforced by the global pandemic outbreak, people's engagements with online platforms underwent notable changes, particularly in search engine trends [10–12], exposures to media reports [13,14], and through quotidian smartphone usages for COVID-19 information [5]. Reliance on the internet has significantly increased due to the overnight change in lifestyles, for example, working and remote learning, imposed by the pandemic on society. The sorts of content consumed, the time and duration spent online, and the purpose of online engagements may be influenced by COVID-19. Furthermore, the digital footprints left by online interactions may reveal information about these changes in user behaviors.

Most importantly, such ubiquitous online footprints may provide useful signals of deteriorating mental health profiles of users during COVID-19. They may capture insights into what was going on in the mind of the user through a non-invasive manner, especially since Google and YouTube Searches are short and succinct and can be quite rich in providing the in the moment cognitive state of a person. On one hand, online engagements can cause fluctuations in mental health. On the other hand, having certain mental health conditions can cause certain types of online behaviors. This opens up possibilities for potential healthcare frameworks that leverage pervasive computing approaches to monitor mental health conditions and deliver interventions on-time.

### Prior Work

Extensive researches have been conducted on a population level, correlating mental health problems with user behaviors on social platforms [15,16], especially among young adolescents. Researchers monitored Twitter to understand mental health profiles of the general population such as suicide ideations [17] and depressions [18]. Similar researches have been done with Reddit, where anxiety [19], suicide ideations [17], and other general disorders were studied [20,21]. Another popular public platform is Facebook, and experiments have been done studying anxiety, depression, body shaming, and stress online [22,23]. However, such studies were limited to macro observations and failed to identify individuals in need of mental health assistance. In addition, it has been shown that college student communities rely heavily on YouTube for both academic and entertainment purposes [24,25]. Yet, abundant usages may lead to compulsive YouTube engagements [26], and researchers have found that social anxiety is associated with YouTube consumptions in a complex way [27].

During COVID-19, multiple studies have reported deteriorating mental health conditions in various communities [1–3,28], such as nation-wise [29,30], across the healthcare industry [31,32], and among existing mental health patients [33]. Besides, online behaviors during COVID-19 have been explored, especially for web searches related to the pandemic [10–12] and abnormal TV consumptions during the lockdown [13]. Many of the behavioral studies also discussed the effects of online interactions on the spread, misinformation, knowledge, and protective measures of COVID-19, including the roles of YouTube [34–36] and other platforms [37]. [38] investigated hate speech targeting the Chinese and Asian communities on Twitter during COVID-19.

Ubiquitous data has been proved to be useful in detecting mental health conditions. Mobile sensor data, such as GPS logs [39,40], electrodermal activity, sleep behavior, motion, and phone usage patterns [41,42] has been applied in investigating depressive symptoms. [43] found that individual private Google Search histories can be used to detect low self-esteem conditions among college students. [5] examined the longitudinal changes in mental health and smartphone usages through ecological momentary assessments (EMAs) during COVID-19 among college populations. Yet, none of the previous studies evaluated the relationship *between* individual online behaviors (Google Search and YouTube) and the deterioration in mental health conditions during COVID-19.

### Goal of This Study
It has been shown that online platforms preserve useful information about the mental health conditions of users, and COVID-19 is jeopardizing the mental well-being of the global community. Thus, we demonstrate the richness of online engagement logs and how it can be leveraged to uncover alarming mental health conditions during COVID-19. In this study, we aim to examine whether the changes in user behaviors during COVID-19 have a relationship with deteriorating mental health profiles. We focus on Google Search and YouTube usages, and we investigate if the behavior shifts when engaging with these two platforms signify worsened mental health conditions. We hypothesize that late-night activities, compulsive and continuous usages, time away from online platforms, porn and news consumptions, and keywords related to health, social engagements, personal affairs, and negative emotions may play a role in deteriorating mental health conditions.

The scope of the study covers undergraduate students in the U.S. We envision this project as a pilot study: it may lay a foundation for mental health surveillance and help delivery frameworks based on pervasive computing and ubiquitous online data. Compared to traditional interviews and surveys, such a non-invasive system may be cheaper, efficient, and avoid being blocked by social stigmas while notifying caregivers on-time about individuals at risk.

## Methods

### Recruitment and Study Design

We recruited a cohort of undergraduate students, all of whom were at least 18 years old and have an active Google account for at least 2 years, from the University of Rochester River Campus, Rochester, NY, U.S.A. Participation was voluntary, and individuals had the option to opt-out of the study at any time, although we did not encounter any such cases. We collected individual-level longitudinal online data (Google Search and YouTube) in the form of private history logs from the participants. For every participant, we measured the depression and anxiety levels via the clinically validated Patient Health Questionnaire-9 (PHQ-9) and Generalized Anxiety Disorder-7 (GAD-7), respectively. Basic demographic information was also recorded. There were in total two rounds of data collection: the first round during January 2020 (prior to the pandemic) and the second round during May 2020 (during the pandemic). During each round, for each participant, the anxiety and depression scores were assessed, and the change in mental health conditions was calculated in the end. The entire individual online history data up untill the date of participation was also collected in both rounds from the participants. Figure 2 gives an illustration of the recruitment timeline and two rounds of data collections. All individuals participated in both rounds and were compensated with 10-dollar Amazon gift cards during each round of participation.

Given the sensitivity and proprietary nature of private Google Search and YouTube histories, we leveraged the Google Takeout web interface [44] to share the data with the research team. Prior to any data cleaning and analysis, all sensitive information such as the name, email, phone number, social security number, and credit card information was automatically removed via the Data Loss Prevention (DLP) API [45] of Google Cloud. For online data and survey response storage, we utilized a HIPAA-compliant cloud-based secure storing pipeline. The whole study design, pipelines, and survey measurements involved were similar to our previous setup in [43] and have been approved by the Institutional Review Board (IRB) of the University of Rochester.

### Online Data Processing and Feature Extractions

The Google Takeout platform enables users to share the entire private history logs associated with their Google accounts, and as long as the account of the user was logged in, all histories would be recorded regardless of which device the individual was using. Each activity in Google Search and YouTube engagement logs were timestamped, signifying when the activity happened to the precision of seconds. Besides, for each Google Search, the history log contained the query text input by the user. It also recorded the URL if the user directly input a website address to the search engine. For each YouTube video watched by the user, the history log contained the URL to the video. If the individual directly searched with keyword(s) on the YouTube platform, the history log also recorded the URL to the search results.

In order to capture the change in online behaviors for the participants, we first introduced a set of features that quantifies certain aspects of how individuals interact with Google Search and YouTube. The set of features was calculated for each participant separately. Individual-level behavior changes were then obtained by examining the variations of the feature between January to mid-March of 2020 (prior to the outbreak) and mid-March to May of 2020 (after the outbreak).

Concretely, we defined 5 features and cut the longitudinal data of each participant into two segments by mid-March, around the time of the COVID-19 outbreak in the U.S and campus lockdown. The two segments spanned 2.5 months before and after mid-March, respectively, and data before January 2020 was discarded. The same feature was extracted from both segments of data, and the change was calculated. Such change was referred to as the behavior shifts during the pandemic and lockdown. Figure 2 gives an illustration of data segmentations and feature development pipelines.

### Late Night Activities

We defined late-night activities (LNA) as the activities happened between 10:00 P.M. and 5:00 A.M. of the next day, regardless of Google Search or YouTube. For each participant, we counted the numbers of late-night activities before ($LNA_{before}$) and after the outbreak ($LNA_{before}$), respectively. We then calculated the percentage change of late-night activities and used it as a behavior shift feature:

*Equation 1. The percentage change of late nigh activities before and after the COVID-19 outbreak.*

$$\% \text{ change of LNA} = \frac{LNA_{after} - LNA_{before}}{LNA_{before}} \quad (1)$$

For the rest of the study, any mentioned percentage or relative changes of features were calculated the same way as above.

### Inactivity Periods

We defined inactivity periods as the periods of time where no Google Search nor YouTube activity was performed. We set a threshold of 7 hours, and we identified all the inactivity periods that were longer than 7 hours for each participant from the online data log. Moreover, we looked at how these inactivity periods were distributed across 24 hours. We obtained the mid-point hour mark for each inactivity period: for example, an inactivity period started at 11 P.M. and ended at 7 A.M. has a mid-point of 3 A.M. With normalization, we received a discrete distribution of inactivity period midpoints over the 24-hour bins. It represented how the time away from Google Search and YouTube of an individual was distributed in a 24-hour period. Such distribution was calculated on the data segments before ($Q_{before}$) and after ($Q_{after}$) the outbreak, respectively. Figure 1 showcases two normalized inactivity midpoint distributions before and after the outbreak: after the outbreak, most of the inactive periods of participant 1 shifted to later hours of the dawn, which was most likely to be a delay in bedtime; for participant 2, the morning inactivity moved earlier, and new inactive periods during the afternoon appeared after the outbreak. One possible explanation could be that participant 2 started to take naps after noon, resulting in midpoints around 5 P.M.

*Figure 1. The normalized inactivity midpoint distributions over 24 hours before and after the outbreak of COVID-19 of two example participants. The threshold for the inactivity is 7 hours. Brighter blocks indicate higher frequencies in that hour.*

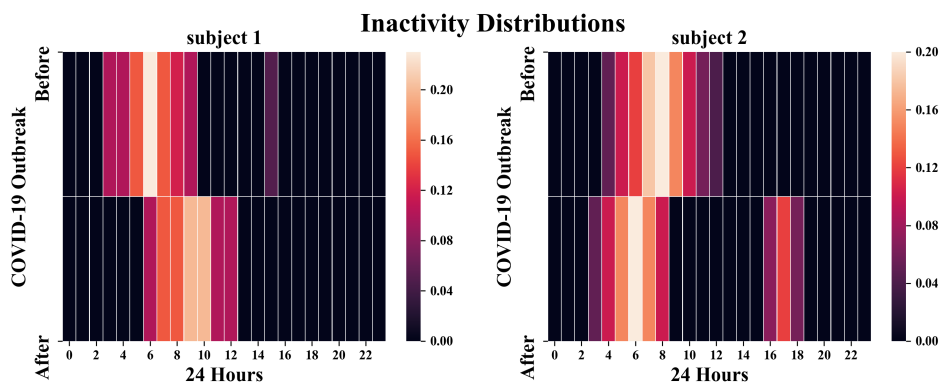

To estimate the difference before and after the outbreak, we calculated the KL-divergence [46] between the two distributions for each participant:

*Equation 2. The KL divergence of inactivity distributions before and after the COVID-19 outbreak.*

$$D_{KL} = \sum_{t=0}^{23} Q_{before}(t) \times \log\left(\frac{Q_{before}(t)}{Q_{after}(t)}\right) \quad (2)$$

The KL-divergence is strictly greater than or equals to 0, and it equals to 0 only when the two distributions are identical.

### Short Event Intervals

We defined a short event interval (SEI) as the period of time that is less than 5 minutes between two adjacent events. It usually occurs when one is consuming several YouTube videos or searching for related content in a roll. We counted the total numbers of short event intervals for each participant before ($SEI_{before}$) and after ($SEI_{after}$) the outbreak, respectively. We calculated the percentage change of SEI the same way as Equation 1 and used it as a behavioral feature.

### LIWC Attributes

The Linguistic Inquiry and Word Count (LIWC) is a toolkit used to analyze various emotions, cognitive processes, social concerns, and psychological dimensions in a given text by counting the numbers of specific words [47]. It has been widely applied in researches involving social media and mental health. For the complete list of linguistic and psychological dimensions LIWC measures, see [47(pp3-4)]. We segmented the data log for each participant by mid-March as two blobs of texts and analyzed the words using LIWC: for Google Search, we input the raw query text; for YouTube, we input the video title. We considered the 'Personal Concerns', 'Negative Emotion', 'Health/illness', and 'Social Words' LIWC dimensions. LIWC categorized words associated with work, leisure, home, money, and religion as 'Personal Concerns'. In the 'Negative Emotion' dimension, LIWC included words related to anxiety, anger, and sadness. Whereas, in the 'Social Words' dimension, LIWC included family, friends, and gender references. The LIWC output the count of words falling in each dimension among the whole text. We quantified the shift in behavior by calculating the percentage change of words in each dimension after the outbreak.

### Google Search and YouTube Categories

We labeled each Google Search query with a category using the Google NLP API [48]. We utilized the official YouTube API to retrieve the information of videos watched by the participants, including the title, duration, number of likes and dislikes, and default YouTube category tags. For a comprehensive list of Google NLP category labels and default YouTube category tags, please refer to [49,50]. There were several categories overlapping with the LIWC dimensions, such as 'Health' and 'Finance', and we regarded the LIWC dimensions as a more well-studied standard. Instead, we focused on the number of activities belonging to the 'Adult' and 'News' categories, which were not presented in the LIWC. We calculated the relative changes of activities in these two categories as the behavior shifts for each participant, the same as Equation 1.

*Figure 2. The study recruitment procedure and feature development process.*

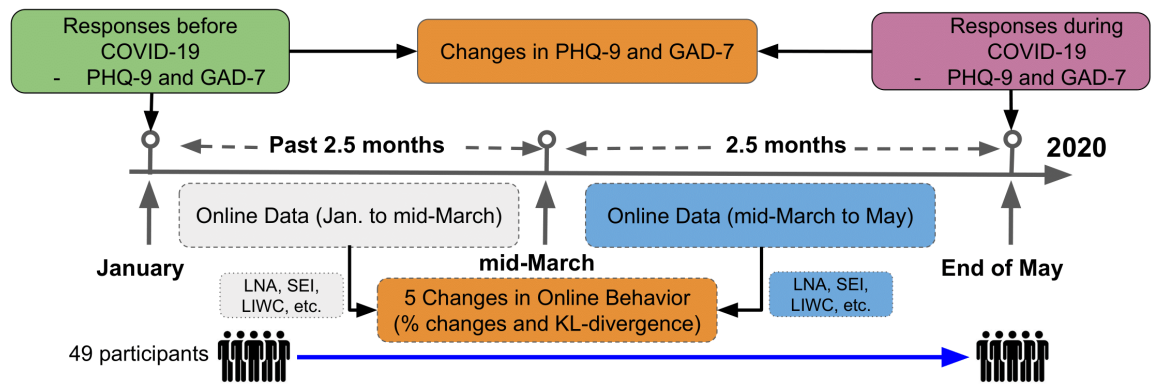

## Measurement Outcomes

### Measurements for Changes in Online Behaviors

There were in total 5 scalar continuous dependent variables measuring various aspects of the changes in online behavior for each participant, as defined above. These variables were extracted from two segments of the online data logs, namely the data before and after the pandemic outbreak. For the *Inactivity Periods*, the measurement was the KL-divergence between inactivity distributions. For the rest 4 behavioral features, the measurements were all in percentage changes.

### Measurements for Mental Health Conditions

For both rounds of the data collection, anxiety levels were assessed using the GAD-7 survey, and depression levels were assessed using the PHQ-9 survey. With two rounds of surveys reported before and after the outbreak, the change in mental health conditions of each participant was obtained. According to [51,52], an increase greater than or equals to 5 in the GAD-7 score may be clinically alarming. Therefore, individuals with an increase ≥ 5 in GAD-7 scores were labeled as the ANX group; the rest were labeled as the non-ANX group. Similarly, as stated in [53], an increase greater than or equals to 5 in the PHQ-9 score may indicate the need for medical interventions. Hence, individuals with an increase ≥ 5 in PHQ-9 scores were labeled as the DEP group; the rest were labeled as the non-DEP group.

### Demographics and Covariates

Besides the online data and mental health surveys, we also collected basic demographic information such as school year, gender, and nationality.

### Statistical Analysis

Before any analysis of mental health conditions, in order to eliminate the possibility of annual confounding factors interfering with the shifts in online behaviors, two-tailed paired independent *t*-tests were performed. We inspected that, in terms of the five quantitative features, whether the online behavior changes happened every year, such as due to seasonal factors, or only during COVID-19 for the whole study population. As mentioned above, we collected the entire Google history log back to the registration date of the Google accounts of all participants. Thus, we computed the online behaviors changes in both 2020 and 2019 for all participants, spanning 2.5 months before and after the mid-March of each year. The behavior changes were dependent between 2020 and 2019 for the same participant. Viewing the cohort as a whole and measured twice, two-tailed paired independent *t*-tests were performed on all 5 behavior features.

For the main experiment, chi-square tests were first performed to investigate the differences in demographics: school year, gender, and nationality. After that, analyses of covariance were conducted to explore the discrepancy between the DEP and non-DEP groups with each of the 5 online behavior features while controlling significant demographic covariates. The same was performed between the ANX and non-ANX groups. Notice that, in this observational study, the independent variable was the binary group, i.e., whether or not the individual had a significant increase in the GAD-7 (or PHQ-9) score. The dependent variables were the 5 behavior changes extracted from the longitudinal individual online data. Experiments were carried out in a one-on-one fashion: anxiety or depression condition was the single independent variable, and one of the 5 online behavior changes was the single dependent variable each time.

Since multiple hypotheses were tested and some dependent variables might be moderately correlated, a Holm's sequential Bonferroni procedure was performed with an original significance level $\alpha$=0.05 to deal with the family-wise error rates.

## Results

### Study Population Statistics

We recruited 49 (N=49) participants in total, and all of them participated in both rounds of the study (response rate=100%). On average, each participant made 2,446 (95% CI 2,120.22-2,481.75) Google Searches and 2,985 (95% CI 2,576.57-3,393.43) YouTube interactions from January to March 14th, and 2578 (95% CI 2,165.16-2,990.96) Google Searches and 3146 (95% CI 2,758.79-3,533.24) YouTube interactions from March 14th to the end of May. Of the 49 participants, 41% (n=20) of them reported an increase in the PHQ-9 score ≥ 5 (the DEP group); 45% (n=22) of them reported an increase in the GAD-7 score ≥ 5 (the ANX group). 37% (n=18) of the participants belonged to the ANX and DEP group simultaneously.

Of the 49 participants, 61% (n=30) of the them were female; 35% (n=17) of the them were male; the rest 4% (n=2) reported non-binary genders. First and second-year students occupied 63% (n=31) of the whole cohort, and the rest were third and fourth-year students (n=24). 80% (n=39) of the participants were U.S. citizens, and the rest (n=10) were international students. A complete breakdown of demographics and group separations are given in Table 1.

*Table 1. Demographics of the study population.*

| Demographic | ANX (n=22) | non-ANX (n=27) | DEP (n=20) | non-DEP (n=29) |
|---|---|---|---|---|
| Female, n (%) | 17 (77) | 13 (48) | 17 (85) | 13 (45) |
| U.S. citizen, n (%) | 17 (77) | 22 (81) | 15 (75) | 24 (83) |
| 1st and 2nd-year, n (%) | 15 (68) | 16 (59) | 13 (65) | 18 (62) |

### Evaluation Outcomes

The distributions of female participants were not well-stratified. 77% (n=17/22) of the ANX group were female while 48% (n=13/27) of the non-ANX group were female; 85% (n=17/20) of the DEP group were female while 45% (n=13/29) of the non-DEP group were female (Table 1). This observation among female students is consistent with the statistics reported in [4]. Chi-square tests showed that being female had a significant difference between the ANX and non-ANX group ($P=.07$, $\chi_1^2=3.2$); it also had a significant difference between the DEP and non-DEP group ($P=.01$, $\chi_1^2=6.4$). Meanwhile, being a U.S. citizen did not show a significant difference in deteriorating anxiety ($P=.99$, $\chi_1^2<0.1$) nor depression ($P=.76$, $\chi_1^2=0.1$); being a lower-class student (first or second-year) did not show a significant difference in deteriorating anxiety ($P=.73$, $\chi_1^2=0.1$) nor depression ($P=.93$, $\chi_1^2<0.1$). Thus, the gender factor was controlled for the rest of the study.

The two-tailed paired independent *t*-tests mentioned at the beginning of Statistical Analysis was designed to rule out seasonal factors in online behavior changes but focus on COVID-19 before any of the main experiments, and they reported *P*<.001 for all 5 quantitative features. Hence, the presence of annual or seasonal factors accountable for online behavior changes was neglectable, and it was safe to carry out the following main experiment. This is consistent with one of the main conclusions in [5] that, when comparing the longitudinal data between different years, behaviors during COVID-19 shifted drastically.

For each group (ANX, non-ANX, DEP, and non-DEP), the average percentage changes in *Late Night Activities*, *Short Event Intervals*, *LIWC Attributes*, and *Google Search and YouTube Categories* were all positive increases.

### *Depression Group Analysis*

Analyses of covariance were performed to investigate the online behavior differences between the DEP and the non-DEP groups, ruling out the gender factor. We dummy-coded the categorical gender factor as a continuous covariate. For *Late Night Activities*, the DEP group (mean=9.70%, 95% CI 8.72%-10.68%) had a higher relative increase than the non-DEP group (mean=7.54%, 95% CI 6.59%-8.49%), and a significant difference was found (*P*=.005, $\eta^2_{partial}$=0.156, $F_{1,46}$=8.53). For *Inactivity Periods*, the DEP group (mean=0.86, 95% CI 0.65-1.06) had a lower divergence, i.e., fewer variations in how the time away from Google products was distributed in a day, than the non-DEP group (mean=1.32, 95% CI 1.20-1.44), and a significant difference was found (*P*<.001, $\eta^2_{partial}$=0.319, $F_{1,46}$=21.55). The DEP group (mean=18.46%, 95% CI 16.51%-20.40%) had more increase in *Short Event Intervals* than the non-DEP group (mean=14.62%, 95% CI 13.28%-15.96%), and a significant difference was found (*P*=.002, $\eta^2_{partial}$=0.183, $F_{1,46}$=10.34).

For the *LIWC Attributes*, the DEP group (mean=9.52%, 95% CI 8.36%-10.78%) had a higher relative increase in 'Personal Concern' keywords than the non-DEP group (mean=7.24%, 95% CI 6.50%-7.98%), and a significant difference was found (*P*=.001, $\eta^2_{partial}$=0.199, $F_{1,46}$=11.45). Similarly, for the prevalence of 'Negative Words' (*P*=.006, $\eta^2_{partial}$=0.153, $F_{1,46}$=8.28), the DEP group (mean=4.26%, 95% CI 3.62%-4.89%) increased more than the non-DEP group (mean=2.97%, 95% CI 2.52%-3.41%); for 'Social Words' (*P*=.006, $\eta^2_{partial}$=0.152, $F_{1,46}$=8.22), the DEP group (mean=3.57%, 95% CI 2.55%-4.58%) increased less than the non-DEP group (mean=5.44%, 95% CI 4.76%-6.12%). 'Health/illness' (*P*=.69, $\eta^2_{partial}$=0.004, $F_{1,46}$=0.17) did not show any significant group difference with a 95% CI. For *Google Search and YouTube Categories*, 'Adult' contents showed a significant difference (*P*=.01, $\eta^2_{partial}$=0.130, $F_{1,46}$=6.85): the DEP group (mean=13.45%, 95% CI 11.23%-15.68%) had a greater increasing consumption than the non-DEP group (mean=9.90%, 95% CI 8.46%-11.35%). 'News' contents did not show any significant group difference (*P*=.39, $\eta^2_{partial}$=0.016, $F_{1,46}$=0.75). Table 2 summarizes these findings in detail. Figure 3 shows the distributions of the percentage increases in online behavior features except for the inactivity divergence in the two groups.

*Table 2. The means, SDs, and statistical test results for the DEP and non-DEP groups.*

| Variables | DEP, mean (SD) | non-DEP, mean (SD) | DEP V.S. non-DEP | | |
|---|---|---|---|---|---|
| | | | *P* value | $\eta^2_{partial}$ | $F_{1,46}$ |
| Late Night Activities (%) | 9.70 (2.04) | 7.54 (2.46) | .005 | 0.156 | 8.53 |
| Inactivity Periods ($D_{KL}$) | 0.86 (0.43) | 1.32 (0.30) | <.001 | 0.319 | 21.55 |
| Short Event Intervals (%) | 18.46 (4.06) | 14.62 (3.46) | .002 | 0.183 | 10.34 |
| **LIWC Attributes (%)** | | | | | |
|     Personal Concern | 9.52 (2.63) | 7.24 (1.91) | .001 | 0.199 | 11.45 |
|     Negative Words | 4.26 (1.33) | 2.97 (1.16) | .006 | 0.153 | 8.28 |
|     Social Words | 3.57 (2.11) | 5.44 (1.77) | .006 | 0.152 | 8.22 |
|     Health/illness | 10.62 (3.96) | 10.60 (4.24) | .69 | 0.004 | 0.17 |
| **Google Search and YouTube Categories (%)** | | | | | |
|     Adult | 13.45 (4.63) | 9.90 (3.73) | .01 | 0.130 | 6.85 |
|     News | 5.16 (3.03) | 5.95 (2.13) | .39 | 0.016 | 0.75 |

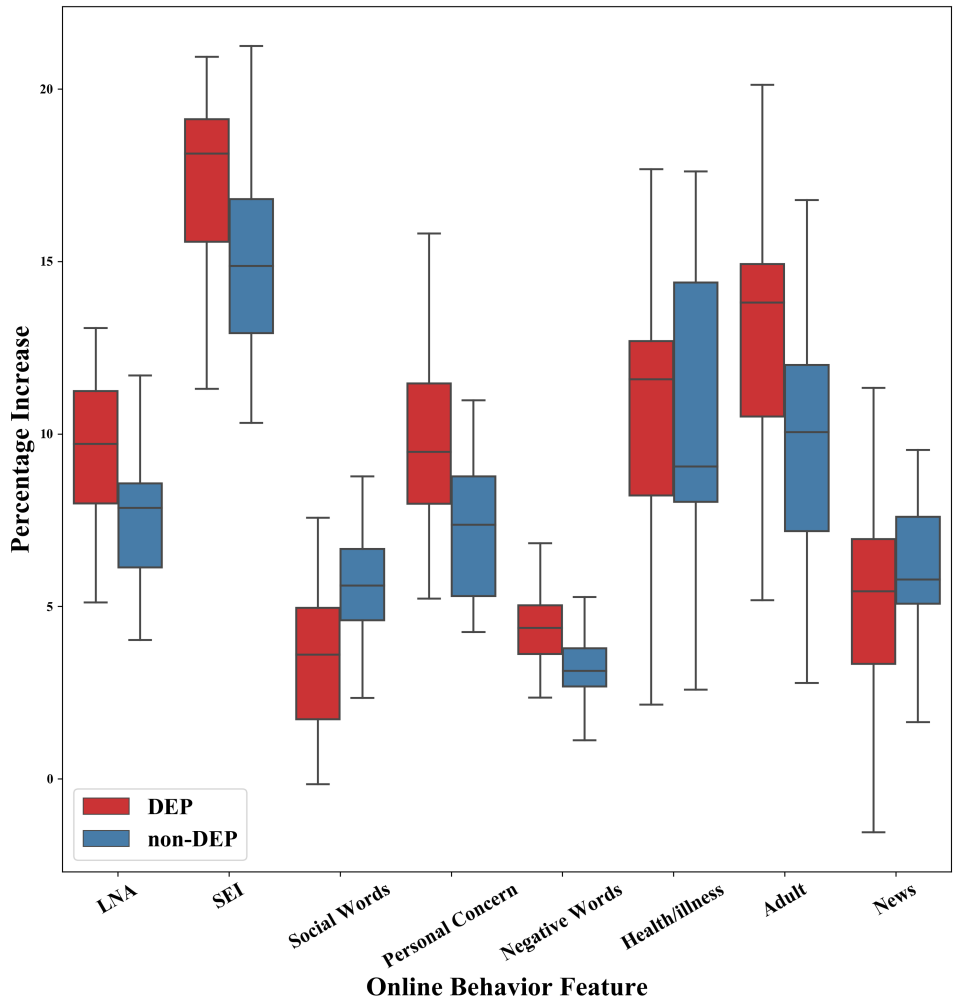

*Figure 3. Comparisons of the 4 online behavior changes measured in percentage increases, except the inactivity distributions, between the DEP (red) and the non-DEP (blue) groups. Boxes cover the 25th and 75th percentiles, and whiskers represent the range of the group. Horizontal solid lines in the boxes represent the medians.*

### Anxiety Group Analysis

Similar trends were found between the ANX and non-ANX groups, partially due to the overlapping with the DEP and non-DEP populations. For *Late Night Activities* ($P=.001$, $\eta^2_{partial}=0.231$, $F_{1,46}=13.85$), the ANX group (mean=9.82%, 95% CI 9.04%-10.61%) had a higher percentage increase than the non-ANX group (mean=7.28%, 95% CI 6.27%-8.29%). For *Inactivity Periods* ($P=.01$, $\eta^2_{partial}=0.135$, $F_{1,46}=7.19$), the ANX group (mean=0.96, 95% CI 0.76-1.16) had a lower divergence, i.e., fewer alterations in the pattern of inactive periods in a 24-hour period, than the non-ANX group (mean=1.27, 95% CI 1.13-1.42). The ANX group (mean=18.09%, 95% CI 16.22%-19.95%) had more increase in *Short Event Intervals* than the non-ANX group (mean=14.64%, 95% CI 13.21%-16.06%), and a significant difference was found ($P=.007$, $\eta^2_{partial}=0.149$, $F_{1,46}=8.05$).

For the *LIWC Attributes*, the ANX group (mean=9.15%, 95% CI 7.88%-10.41%) had a higher relative increase in 'Personal Concern' keywords than the non-ANX group

(mean=7.37%, 95% CI 6.61%-8.14%), and this difference was statistically significant ($P$=.02, $\eta^2_{partial}$=0.115, $F_{1,46}$=5.99). We found a similar result for 'Negative Words' ($P$=.01, $\eta^2_{partial}$=0.125, $F_{1,46}$=6.59) where the ANX group (mean=4.12%, 95% CI 3.48%-4.76%) had higher usages than the non-ANX group (mean=2.98%, 95% CI 2.53%-3.43%). 'Health/illness' ($P$=.52, $\eta^2_{partial}$=0.009, $F_{1,46}$=0.42) and 'Social Words' ($P$=.10, $\eta^2_{partial}$=0.057, $F_{1,46}$=2.77) did not show any significant group difference with a 95% CI. For *Google Search and YouTube Categories*, neither 'Adult' ($P$=.25, $\eta^2_{partial}$=0.028, $F_{1,46}$=1.33) nor 'News' ($P$=.71, $\eta^2_{partial}$=0.003, $F_{1,46}$=0.14) content showed any significant group difference. For more details, see Table 3. Figure 4 shows the distributions of the percentage increases in online behavior features except for the inactivity divergence in the two groups.

*Table 3. The means, SDs, and statistical test results for the ANX and non-ANX groups.*

| Variables | ANX, mean (SD) | non-ANX, mean (SD) | ANX V.S. non-ANX | | |
|---|---|---|---|---|---|
| | | | $P$ value | $\eta^2_{partial}$ | $F_{1,46}$ |
| Late Night Activities (%) | 9.82 (1.73) | 7.28 (2.51) | .001 | 0.231 | 13.85 |
| Inactivity Periods ($D_{KL}$) | 0.96 (0.45) | 1.27 (0.35) | .01 | 0.135 | 7.19 |
| Short Event Intervals (%) | 18.09 (4.11) | 14.64 (3.52) | .007 | 0.149 | 8.05 |
| **LIWC Attributes (%)** | | | | | |
| Personal Concern | 9.15 (2.78) | 7.37 (1.89) | .02 | 0.115 | 5.99 |
| Negative Words | 4.12 (1.42) | 2.98 (1.12) | .01 | 0.125 | 6.59 |
| Social Words | 4.00 (2.46) | 5.22 (1.61) | .10 | 0.057 | 2.77 |
| Health/illness | 10.83 (4.26) | 10.44 (3.91) | .52 | 0.009 | 0.42 |
| **Google Search and YouTube Categories (%)** | | | | | |
| Adult | 12.37 (5.00) | 10.52 (3.80) | .25 | 0.028 | 1.33 |
| News | 5.40 (2.98) | 5.81 (2.15) | .71 | 0.003 | 0.14 |

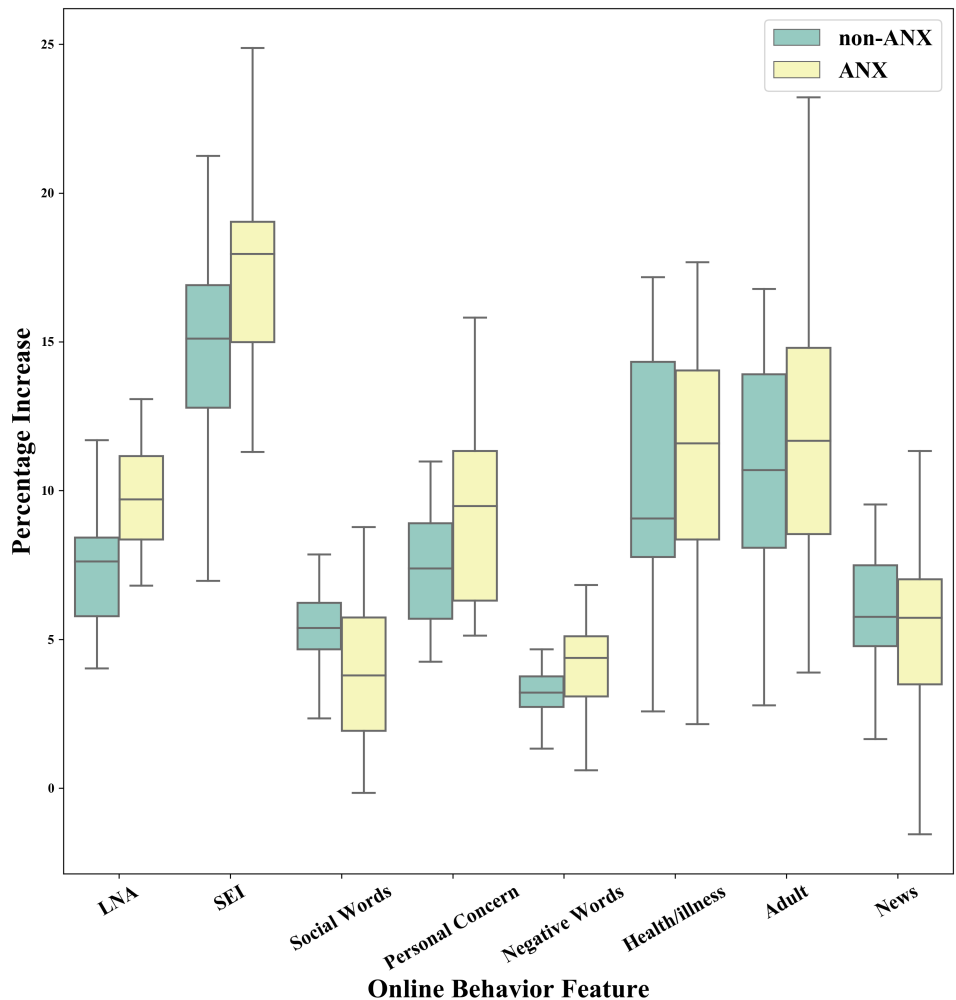

*Figure 4. Comparisons of the 4 online behavior changes measured in percentage increases, except the inactivity distributions, between the ANX (yellow) and the non-ANX (green) groups. Boxes cover the 25th and 75th percentiles, and whiskers represent the range of the group. Horizontal solid lines in the boxes represent the medians.*

## Discussion

In this study, we collected longitudinal individual-level Google Search and YouTube data from college students, and we measured their anxiety (GAD-7) and depression (PHQ-9) levels before and after the outbreak of COVID-19. We then developed explainable features from the online data logs and quantified the online behavior shifts of the participants during the pandemic. We also calculated the change in mental health conditions for all participants. Our experiment examined the differences between groups with and without deteriorating mental health profiles in terms of these online behavior features. To the best of our knowledge, we are the first to conduct observational studies on how mental health problems and Google Search and YouTube usages of college students are related during COVID-19.

### Principal Results

Our results showed significant differences between groups of college students with and without worsened mental health profiles in terms of online behavior changes during the pandemic. The features we developed based on online activities were all explainable and preserved certain levels of interpretability. For example, the *Short Event Intervals* and *Inactivity Periods* measured the consecutive usages and time away from Google Search and YouTube, which were inspired by previous studies on excessive YouTube usages [26], internet addictions [54], and positive associations with social anxiety among college students [27]. Our results indicated that individuals with meaningful increasing anxiety or depressive disorders during the pandemic tended to have long usage sessions (multiple consecutive activities with short time intervals) when engaging with Google Search and YouTube.

Moreover, ANX and DEP individuals tended to maintain their regular time-away-from-internet patterns regardless of the lockdown as the KL-divergence was low. One possible reason could be that depressed people tend to spend more time at home as regular lifestyles [40,55], and thereby, after the lockdown, the living environment did not alter much. We further found that the majority of the inactivity periods longer than 7 hours had midpoints around 5 to 6 A.M. for all individuals, which were most likely to be the sleeping period. Well-established previous researches stated that depressed individuals have more disrupted sleeping patterns and less circadian lifestyles [40,56,57], but they are not validated for special periods such as COVID-19. We instead focused on comparing the distributions of time away from Google before and after the outbreak of COVID-19, and we had an emphasis on the behavior *changes* of groups with and without worsened mental disorders.

Besides, the increase in *Late Night Activities* corresponded with previous studies in sleep deprivation and subsequent positive correlations with mental health deteriorations [58,59]. Our results demonstrated that individuals with significant worsened anxiety or depressive symptoms during the pandemic were indeed likely to stay up late and engage more online. The above three features captured the temporal aspects of user online behaviors, and they have shown statistically significant differences between groups.

Additionally, our analysis found that there was a significant difference in the amount of adult and porn consumption between individuals with and without worsening depression, which adheres to previous findings that people suffering from depression and loneliness are likely to consume more pornographies [60,61]. For the LIWC features, 'Personal Concern' and 'Negative Emotion' keywords appeared more frequently among students in the ANX group, and previous research showed that negative YouTube videos tended to receive more attention from vulnerable individuals [62]. For the DEP group, 'Social Words' became less prevalent than the non-DEP group. This was consistent with studies on patterns of social withdrawal and depression [40,63,64], and social interactions and isolations have been recognized by [65] as one of the priorities in mental illness prevention, especially during COVID-19 [30]. These attributes captured the semantic aspect of user online behaviors. The prevalence of personal affair, social activity, and negative keywords as well as porn consumption have shown statistically significant differences between groups.

Many researchers have reported that there has been a significant boost in health and news-related topics, at the population level, in various online platforms during COVID-19. This is partly due to additional measures taken by individuals, various stakeholders, and agencies with regards to preventive measures [11,35,36], daily statistics [10,12,13], and healthcare (mis)information [34,36,37], However, unlike many, our investigation was carried out considering individual-level Google Search and YouTube engagement logs, and our analysis did not reveal any significant spikes in 'News' and 'Health/illness' category between the groups of individuals with deteriorating anxiety and depression during the pandemic. One possible explanation for such observation can be due to the target population (college students) of our study who may prefer to follow news from other popular platforms such as social media.

Finally, COVID-19 has shaken the foundation of human society and forced us to alter daily lifestyles. The world was not ready for such a viral outbreak. Since there is no cure for COVID-19, it, or an even more deadly viral disease, may resurface at different capacities in the near future. Society may be forced to rely on technologies even more and employ remote learning, working, and socializing for a longer period of time. It is important that we learn from our experience of living through the initial COVID-19 outbreak and take necessary measures to uncover the changes in online behaviors, investigating how that can be leveraged to understand and monitor various mental health conditions of individuals in the least invasive manner. Furthermore, we hope our work paves the path for technology stakeholders to consider incorporating various mental health assessment monitoring systems using user engagements, following users' consent in a privacy-preserving manner. They can periodically share the mental health monitoring assessment report with respective users based on their online activities, education, and informing users about their current mental health. This can eventually encourage individuals to acknowledge the importance of mental health and take better care of themselves.

## Limitations

First, while most of the online behavioral features we developed showed significant differences between groups of students with and without deteriorating anxiety and depressive disorders during COVID-19, our study cohort only represented a small portion of the whole population suffering from mental health difficulties. Therefore, further studies are required to investigate if the significant behavioral changes still hold among more general communities, not limiting to college students. Nonetheless, we argue that the explainable features we constructed, such as late-night activities, continuous usages, inactivity, pornography, and certain keywords, can remain behaviorally representative and be applied universally across experiments exploring the relationship between mental health and online activities during the pandemic.

Second, in this work, we studied the relationship between user online behaviors and the fluctuations in mental health conditions during COVID-19. Any causal relationship between online behavior and mental disorders is beyond the scope of this work. As one can readily imagine, online behavioral changes could both contribute to or be caused by deteriorating anxiety or depressive disorders. Moreover, though we included preliminary demographic information as covariates, there remains the possibility of other confounding factors. In fact, both the shifts in online behaviors and deteriorating mental health profiles may be due to common factors such as living conditions, financial difficulties, and other health problems during the pandemic. Nor there was any causal direction implied between COVID-19 and online behavior changes, which was introduced in the first paragraph of Statistical Analysis as a precaution before the main experiments.

### Ethical and Privacy Concerns

Albeit a pilot study, our results indicated that it is possible to build an anxiety and depression surveillance system based on passively collected private Google data histories during COVID-19. Such non-invasive systems shall be subject to rigorous data security and anonymity checks. Necessary measures need to be in place to ensure personal safety and privacy concerns when collecting sensitive and proprietary data such as Google Search logs and YouTube histories. Even in pilot studies, participants shall preserve full rights over their data: they may choose to opt-out of the study at any stage and remove any data shared in the system.

Moreover, anonymity and systematic bias elimination shall be enforced. As an automatic medical screening system based on pervasive data, it has been extensively studied that such frameworks are prone to implicit machine learning bias during data collection or training phases [66–68]. Black-box methods should be avoided as they are known to be vulnerable to adversarial attacks and produce unexplainable distributional representations [69,70]. Anonymizing data and obscuring identity information should be the first step in data debiasing.

In the end, to what extent should caregivers trust a clinical decision made by machines remains an open question. We believe that possible pervasive computing frameworks shall play the role of a smart assistant, at most, to the care providers. Any final intervention or help delivery decision should be made by healthcare professionals who understand both the mental health problems and the limitations of automatic detection systems in clinical settings.

### Acknowledgments
This research was supported in part by grant W911NF-15-1-0542 and W911NF-19-1-0029 with the US Defense Advanced Research Projects Agency (DARPA) and the Army Research Office (ARO). We acknowledge the contributions by Michael Giardino, Adira Blumenthal, and Ariel Hirschhorn at the beginning phase of the project.

### Conflicts of Interest
Disclose any personal financial interests related to the subject matters discussed in the manuscript here. For example, authors who are owners or employees of Internet companies that market the services described in the manuscript will be disclosed here. If none, indicate with "none declared".

### Abbreviations

ANX: the group of participants reported an increase in the GAD-7 score ≥ 5
COVID-19: coronavirus disease
DEP: the group of participants reported an increase in the PHQ-9 score ≥ 5
EMA: ecological momentary assessment
LIWC: Linguistic Inquiry and Word Count
LNA: Late night activity
non-ANX: the group of participants did NOT report an increase in the GAD-7 score ≥ 5
non-DEP: the group of participants did NOT report an increase in the PHQ-9 score ≥ 5
SEI: Short event interval